\RequirePackage{fix-cm}
\documentclass[twocolumn,epjc3,final]{svjour3}  
\smartqed  
 \usepackage{mathrsfs}
\usepackage{amsmath, txfonts}
\usepackage{graphicx,bm,booktabs,bm}
\usepackage{longtable,lscape}
\usepackage{overpic,multirow,color}

\definecolor{cover}{rgb}{0.77,0.87,0.88}
\definecolor{blueone}{rgb}{0.1,0.1,.7}
\definecolor{citec}{rgb}{0.14,0.47,0.09}
\definecolor{two}{rgb}{0.0,0.5,0.}
\definecolor{three}{rgb}{.5,.1,0.15}
\usepackage[bookmarks=true,bookmarksopen=false,plainpages=false,breaklinks=true,
   bookmarksnumbered=true,hypertexnames=false,
   filecolor=blue,urlcolor=three,menucolor=three,
   linkcolor=three,citecolor=blueone, colorlinks,
   anchorcolor=blue,runcolor=pink,frenchlinks=red
   pdfstartview=FitH,pdftitle=title,%
   pdfauthor=author]{hyperref}
              \allowdisplaybreaks[4]

\graphicspath{{Figures/}} %

\journalname{Eur. Phys. J. C}
\begin{document}
\title{Spectrum of the $uudc \bar c$ hidden charm pentaquark with an SU(4) flavor-spin hyperfine interaction}

\author{Fl. Stancu\thanksref{e1,addr1}
}                     
\thankstext{e1}{e-mail: fstancu@ulg.ac.be}

\institute{Universit\'e de Li\`ege, Institut de Physique, B.5, Sart Tilman B-4000 Li\`ege 1, Belgium\label{addr1}
}

\date{Received: date / Revised version: date}
%
\maketitle

\begin{abstract}
We study a few of the lowest states of the pentaquark $uudc\overline{c}$,
of positive and  negative parity,  
in a constituent quark model with an SU(4) flavor-spin hyperfine interaction. 
For  positive parity we introduce space wave functions of appropriate permutation symmetry 
with one unit of orbital angular momentum in the internal motion of the four-quark subsystem
or an orbital excitation between the antiquark and the four quark subsystem which remains 
in the ground state.
We show that the lowest positive parity states $1/2^+, 3/2^+$ are provided by the first 
alternative and are located below the $1/2^-$ and the $1/2^+$ states 
with all quarks in the ground state. We compare our 
results with the LHCb three narrow pentaquark structures reported in 2019.
\end{abstract}

%

\section{Introduction}

A considerable wave of interest has been raised by the 2015 observation  
of the LHCb collaboration of two resonances  $P^+_c(4380)$ ($\Gamma = 205 \pm 18 \pm 86$ MeV)
and
$P^+_c(4450)$ ($\Gamma = 39 \pm 5 \pm 19$ MeV) in the  $\Lambda^0_b \rightarrow J/\psi K^- p$ decay, 
interpreted as hidden charm pentaquarks of structure $uud c \bar c$ \cite{Aaij:2015tga}. 
The preferred quantum numbers, which gave the best solution, were $3/2^-, 5/2^+$.
However, the quantum numbers $3/2^+$, $5/2^-$ and  $5/2^-$,  $3/2^+$ were also
acceptable. Various possible interpretations of the 2015 LHCb resonances as kinematical effects, molecular states or
compact pentaquarks were reviewed, see for example Refs. \cite{Ali:2017jda}, \cite{Karliner:2017qhf}
and \cite{Liu:2019zoy}.

Recently the LHCb Collaboration have updated their analysis of the $\Lambda^0_b \rightarrow J/\psi K^- p$ decay
\cite{Skwarnicki,Aaij:2019vzc}  
and observed three narrow peaks with masses and widths as given in Table  \ref{lhcb2019},
where the entirely new $P^+_c(4312)$ has a $7.3 \sigma $ statistical significance and the other two resonances replace the previous $P^+_c(4450)$
now resolved at $5.4 \sigma $ significance. 
The  broad $P^+_c(4380)$ resonance observed in 2015 awaits confirmation and  a proper 
identification of the spin-parity  is necessary in all cases. 

\renewcommand\tabcolsep{0.470cm}
\begin{table}[h!]
\caption{Masses and decay widths of the 2019 LHCb resonances \cite{Skwarnicki,Aaij:2019vzc}.  
\label{lhcb2019}}
\begin{tabular}{ccccccccccccccc}\hline
 Resonance  &  Mass  (MeV)  &  Width  (MeV)   \\
\hline
$P^+_c(4312)$  & 4311.9$\pm$ 0.7 $^{+6.8}_{-0.6}$ & 9.8$\pm$ 2.7 $^{+3.7}_{-4.5}$ \\[0.5ex]
$P^+_c(4440)$  & 4440.3$\pm$ 1.3 $^{+4.1}_{-4.7}$ &20.6$\pm$ 4.9 $^{+8.7}_{-10.1}$ \\[0.5ex]
$P^+_c(4457)$  & 4457.3$\pm$ 0.6 $^{+4.1}_{-1.7}$ & 6.4$\pm$ 2.0 $^{+5.7}_{-1.9}$\\[0.5ex]
\hline
\end{tabular}
\end{table}

The mass of $P^+_c(4312)$ lies a few MeV below the $\Sigma^+_c \overline{D}^0$
threshold  (4318 MeV)
and the masses of $P^+_c(4440)$ and $P^+_c(4457)$ lie a few MeV below the $\Sigma^+_c \overline{D}^{*0}$ threshold (4460 MeV).
Proximity of the 
 $\Sigma^+_c \overline{D}^0$ and  $\Sigma^+_c \overline{D}^{*0}$
thresholds to these narrow peaks made 
their interpretation as molecular S-wave of $\Sigma^+_c \overline{D}^0$ 
or $\Sigma^+_c \overline{D}^{*0}$   quite natural  \cite{Guo:2019kdc,Guo:2019fdo,Xiao:2019mst,Xiao:2019aya,Shimizu:2019ptd,Cheng:2019obk}. 
In such an interpretation,  they all acquire a negative parity. In Ref. \cite{Burns:2019iih} it is shown that the usual molecular scenarios 
cannot explain the production rate of $P_c$ states in $\Lambda_b$ decays.
Moreover, the paper points out the need to couple the $\Sigma^+_c \overline{D}^{*0}$ and the $\Lambda_c(2595) \overline{D}$ channels
due to the very close proximity of their thresholds. As a result, the second and the third  resonances of Table \ref{lhcb2019}
acquire opposite parities,
namely   $J^P(4440) = {\frac{3}{2}}^-$  and   $J^P(4457) = {\frac{1}{2}}^+$ and respectively.


The 2019 LHCb pentaquarks have also been analysed in the compact diquark model  \cite{Ali:2019npk}.
This analysis has been extended in Ref. \cite{Ali:2019clg}, and presently includes
a spin-spin, a spin-orbit and a tensor interaction, obtained from the integration
in the color space of the one gluon exchange (OGE) interaction between quarks \cite{DeRujula:1975qlm}.
The basis contains only the color $\bar 3$ configuration, therefore is smaller than the basis used in the 
present paper. Also there is a belief that the spin 0 diquarks are more 
tightly bound than the spin 1 diquarks. The lowest positive and negative parity states were calculated and it was found that 
the lowest state has negative parity.

Here we explore the spectrum of pentaquarks within a quark model 
\cite{Glozman:1995fu,Robson:1988mp,Glozman:1997ag}, which has  a flavor dependent hyperfine interaction.  Contrary to the 
OGE model result, the lightest pentaquark  has positive parity, as shown below.   
As a general feature, the chromomagnetic (color-spin) and the flavor-spin interactions predict contradictory results regarding the parity
of open charm
exotic systems, as the tetraquark $uu \bar c \bar c$, the pentaquarks $uudd \bar c$,
$uuds \bar c$ and the hexaquark $uuddsc$ \cite{Stancu:1998pr}.

In  Refs. \cite{Glozman:1995fu,Glozman:1997ag} 
the hyperfine splitting in
hadrons is obtained from the short-range part of the Goldstone boson exchange
interaction between quarks. The main merit of the model   
is that it reproduces the correct ordering of positive and negative parity
states  in the baryons spectrum   in
contrast to the OGE model. On the other hand it does not apply to the hyperfine
splitting in mesons because
it does not explicitly contain a quark-antiquark
interaction.

Here we extend   
the  model of Refs. \cite{Glozman:1995fu,Glozman:1997ag} from SU(3) to SU(4) in order to incorporate the
charm quark. The extension is made in the spirit of the phenomenological approach
of Ref. \cite{Glozman:1995xy} where, in addition to Goldstone bosons of 
the hidden approximate chiral symmetry of QCD,  the flavor exchange interaction is augmented  
by additional exchange of $D$ mesons between $u, d$ and $c$ quarks and 
of $D_s$ mesons between $s$ and $c$ quarks. The model provided a satisfactory description of 
the heavy flavor baryons known at that time. With the SU(4) extension we give a more fundamental
ground to the work of Ref. \cite{Glozman:1995fu} and test the corresponding approximations.
In the following we shall use the name of flavor-spin (FS) model for the SU(4) extension. 
The extended flavor exchange allows a complete antisymmetrization of quarks even 
for quarks of widely different mass, since the Young diagrams (or partitions) serve to label the irreducible 
representations both of the permutation and of the SU(4) group \cite{Stancu:1991rc}.

Presently we study  the pentaquarks of structure $uud c \bar c$
in order to see whether or not the FS model can accommodate the new
resonances observed at LHCb  \cite{Aaij:2019vzc}. The approach is similar to that previously used 
in Ref. \cite{Stancu:1998sm} for the positive parity  
$uudd \bar c$ and $uuds \bar c$ pentaquarks. That work preceded the H1 experiment 
\cite{Aktas:2004qf}
which has found evidence,
so far not confirmed by other experiments, for a narrow baryon resonance
in the $D^{*\pm} p$ invariant mass, interpreted as a $uudd \bar c$ pentaquark.
The mass prediction made  in Ref. \cite{Stancu:1998sm} is not far from
the observation of the H1 experiment.

The parity of the
pentaquark is given by P$\ = {\left({-}\right)}^{{\ell\ +\ 1}}$. For the lowest positive parity
states (see below) the subsystem
of four quarks is defined to carry an angular momentum  $\ell$ = 1 in their internal motion.
This implies that this subsystem must be in a state of
orbital symmetry ${\left[{31}\right]}_{O}$. Although the kinetic energy of such
a state is higher than that of the totally symmetric ${\left[{4}\right]}_{O}$
state, a crude estimate 
based on the simplified interaction  (\ref{schematic}), suggests that the
${\left[{31}\right]}_{O}$ symmetry leads to a stable pentaquark of positive parity
and  the lowest state of symmetry ${\left[{4}\right]}_{O}$ is unbound.

In the exact SU(4) limit, like in the exact SU(3) limit,  the flavor-spin interaction introduced in the next section 
takes the following form
\begin{equation}\label{schematic}
{V}_{\chi }\ =\ -\ {C}_{\chi }\ \sum\limits_{ i\ <\ j}^{} {\lambda }_{
i}^{ F} \cdot {\lambda }_{ j}^{ F} \ {\bm {\sigma }}_{i} \cdot {\bm {\sigma
}}_{j}
\end{equation}
with $\lambda_i^F$ the Gell-Mann matrices,  $\bm{\sigma}_i$ the Pauli matrices  and $C_{\chi}$
an equal strength constant for all pairs.

\renewcommand\tabcolsep{0.03cm}
\begin{table}
\caption[matrix]{The simplified interaction - $ V_{\chi} $/$C_{\chi} $, Eq. (\ref{schematic}), 
integrated in the flavor-spin ($FS$) space for   four quark subsystems
in three different states defined in column 2 in the $FS$ coupling. The notation is:
$O$ for orbital, $C$ for color, $F$ for flavor, $S$ for spin. }
 \label{exactlimit}
\begin{tabular}{ccccccccccccccc}
\hline
    State & Definition &  L &  S  &  $ \langle\sum\limits_{ i\ <\ j}^{} {\lambda }_{
i}^{ F}.{\lambda }_{ j}^{ F} \ {\vec{\sigma }}_{i}.{\vec{\sigma
}}_{j} \rangle $ \\
\hline
$|1 \rangle$ &  $\left({{\left[{31}\right]}_{O}{\left[{211}\right]}_{C}{\left[{{1}^{4}}\right]}_{
OC}\
;\
{\left[{22}\right]}_{F}{\left[{22}\right]}_{S}{\left[{4}\right]}_{FS}}\right)$
  &  1  &  0  & \hspace{0.5cm} 27  \\[0.5ex]
$|2 \rangle$ & $\left({{\left[{31}\right]}_{O}{\left[{211}\right]}_{C}{\left[{{1}^{4}}\right]}_{
OC}\
;\
{\left[{31}\right]}_{F}{\left[{31}\right]}_{S}{\left[{4}\right]}_{FS}}\right)$
  & 1  &  1  & \hspace{0.5cm} 21   \\[0.5ex]
$|3 \rangle$ &  $\left({{\left[{4}\right]}_{O}{\left[{211}\right]}_{C}{\left[211]\right]}_{
OC}\
;\
{\left[{211}\right]}_{F}{\left[{22}\right]}_{S}{\left[{31}\right]}_{FS}}\right)$
  &  0  &  0  & \hspace{0.5cm} 15   \\[0.5ex]
\hline
\end{tabular}
\end{table}

Let us  consider two  totally antisymmetric states with the orbital symmetry 
${\left[{31}\right]}_{O}$, allowed by Pauli principle. They are
written in the flavor-spin (FS) coupling
scheme, as given in Table \ref{exactlimit}.  They are consistent with the isospin and
$J^P$ combinations of Table 2 of Ref. \cite{Burns:2015dwa}.

The expectation value of (1), as shown in Table \ref{exactlimit} (see the next section for the derivation
of the last column), 
is $-27 \ C_{\chi}$ for
$\left|{\left.{1}\right\rangle}\right.$ and $-21 \ C_{\chi}$ for
$\left|{\left.{2}\right\rangle}\right.$.

First we consider 
the lowest state, i.e. $\left|{\left.{1}\right\rangle}\right.$. The
quark-antiquark interaction is neglected in the description of mesons as well,
so that the meson Hamiltonian contains a
kinetic and a confinement term only. 
If unstable against strong decays, the pentaquark would spilt into a baryon $q^3$
and a meson $Q\overline{Q}$, where $Q$ stands for a heavy flavor. 
Then, in discussing the stability we introduce the quantity
\begin{equation}\label{DeltaE}
\Delta E =\ E(qqqQ \overline{Q}) -\ E(q^3) -\ E(Q\overline{Q})
\end{equation}
where $E$ defines the energy (mass) of a system with the content defined in the bracket. 
In our schematic estimate, we suppose that the confinement energy roughly
cancels out in $\Delta E$. Then, the kinetic energy contribution \, 
 to \,
$\Delta E$ \, is \, $\Delta  KE\ =\ 5/4\ \hbar \omega$ in a harmonic 
oscillator model \cite{HO}.

Table  \ref{exactlimit} shows that 
the FS contribution to the state $\left|{\left.{1}\right\rangle}\right.$ is
$- 27 \ C_{\chi}$. For the nucleon or $\Lambda_c$ it is - 27/2 $ C_{\chi}$  (see next section).
Then we have $\Delta V_{\chi} =\ - 27/2 \ C_{\chi}$.
With $\hbar\omega \approx$ 157 MeV and $C_{\chi}\approx$ 20 MeV \cite{Glozman:1995xy}
this  gives
\begin{equation}\label{state1}
\Delta  E\ =\ {\frac{5}{4}}\ \hbar \omega  \ -\ 27/2\ {C}_{\chi }\ \approx \ - 74\ \mbox{MeV},
\end{equation}
i.e. a considerable binding. This is to be contrasted with the negative parity
pentaquarks similar to those containing light quarks, studied in  Ref. \cite{Genovese:1997tm}. For 
the lowest negative parity state $\left|{\left.{3}\right\rangle}\right.$ of $uud c \bar c $ one  gets
\begin{equation}\label{state3}
\Delta  E\ =\ 3/4\ \hbar \omega\ -\ 3/2\ {C}_{\chi }\ \approx \ 88\ \mbox{MeV},
\end{equation}
i.e. instability. 
Thus the FS
interaction overcomes the excess of kinetic contribution in
${\left[{31}\right]}_{O}$ and generates a lower expectation value for
${\left[{31}\right]}_{O}$ than for ${\left[{4}\right]}_{O}$.

The paper is organized as follows. In Sec. \ref{Hamiltonian} we introduce the 
model Hamiltonian and generalize the two-body matrix elements of the FS interaction from
SU(3) to SU(4). Sec. \ref{coord} describes the orbital part of the 
four quark subsystem constructed to be translationally invariant both for positive 
and negative parity states. Sec. \ref{analytic} summarizes a few analytic details.
Sec. \ref{numerical} exhibits the numerical results for masses and distances
between quarks/antiquarks. A qualitative discussion of the  decay widths is given in Section \ref{width}.
A comparison with previous studies of hidden charm pentaquark based on various versions of the FS model
is made in Sec. \ref{comp}. In the last Section we draw some
conclusions. \ref{lambda} gives the matrices $\lambda^F$ of SU(4). 
\ref{Casimir}  is a reminder of useful group theory formulae for SU(n).
\ref{baryons} exhibits the variational solution for the baryon masses relevant for the 
present study.
In 
\ref{flavor4} we construct the explicit form of flavor states 
of content $uudc$ in the Young Yamanouchi basis for specific irreducible representations $[{f}]_F$. 

\vspace{1cm}

\section{The Hamiltonian}\label{Hamiltonian}
The nonrelativistic FS Hamiltonian has the general form \cite{Glozman:1995fu}
\begin{eqnarray}
H &=& \sum_i m_i + \sum_i\frac{{\bm p}_{i}^2}{2m_i} 
- \frac {(\sum_i {\bm p}_{i})^2}{2\sum_i m_i} + \sum_{i<j} V_{\text{conf}}(r_{ij}) \nonumber\\
&+& \sum_{i<j} V_\chi(r_{ij}),
\label{ham}
\end{eqnarray}
with $m_i$ and ${\bm p}_{i}$ 
denoting the quark masses and momenta respectively
and $r_{ij}$ the distance between the interacting quarks $i$ and $j$. 
The Hamiltonian contains the internal kinetic energy and the linear confining interaction 
\begin{equation}
 V_{\text{conf}}(r_{ij}) = -\frac{3}{8}\lambda_{i}^{c}\cdot\lambda_{j}^{c} \, C
\, r_{ij} \, .
\label{conf}
\end{equation}
The hyperfine part $V_\chi(r_{ij})$ has a flavor-spin structure which presently is extended to SU(4),  
similarly to Refs. \cite{Pepin:1996id} and  \cite{Glozman:1995xy}, except that
in the latter, the term   $ V_{\eta'} \lambda^0_i \cdot \lambda^0_j$ of Eq. (\ref{VCHI}) has been ignored. 
One has
\begin{eqnarray}
V_\chi(r_{ij})
&=&
\left\{\sum_{F=1}^3 V_{\pi}(r_{ij}) \lambda_i^F \lambda_j^F \right. +  \sum_{F=4}^7 V_{K}(r_{ij}) \lambda_i^F \lambda_j^F 
\nonumber \\
&+& \left.  V_{\eta}(r_{ij}) \lambda_i^8 \lambda_j^8 
+V_{\eta^{\prime}}(r_{ij}) \lambda_i^0 \lambda_j^0\right. 
\nonumber \\
&+& \left. \sum_{F=9}^{12} V_{D}(r_{ij}) \lambda_i^F \lambda_j^F\right.     
+ \left. \sum_{F=13}^{14} V_{D_s}(r_{ij}) \lambda_i^F \lambda_j^F \right.
\nonumber \\
&+& \left. V_{\eta_c}(r_{ij}) \lambda_i^{15} \lambda_j^{15} \right\}
\vec\sigma_i\cdot\vec\sigma_j, 
\label{VCHI}
\end{eqnarray}
\noindent
with $\lambda^F_i$ ($F$ = 1,2,...,15) defined in \ref{lambda}
and
$\lambda^0_i = \sqrt{2/3}~{\bf 1}$, where $\bf 1$ is the $4\times4$ unit
matrix. 
Note that the norm of $\lambda^0_i$ is not that satisfying the general property
tr$(\lambda^a \lambda^b)$ = 2 $\delta^{ab}$ if $\bf 1$ is the $4\times4$ unit
matrix \cite{Stancu:1998sm}, but we maintain the norm from  the SU(3)  version of the flavor-spin interaction
in order to use the same coupling constant for $\eta$ and $\eta '$, which gave a good spectrum
for strange baryons.
At this stage it is useful to mention the study of Ref. \cite{Shuryak:1997xd}
where it has been shown that $\eta'$ has a significant charm component.

In the SU(4) version
the interaction (\ref{VCHI})
contains $\gamma = \pi, K, \eta, D, D_s, \eta_c$ and $\eta '$
meson-exchange terms.  Every $V_{\gamma} (r_{ij})$ is
a sum of two distinct contributions: a Yukawa-type potential containing
the mass of the exchanged meson and a short-range contribution of opposite
sign, the role of which is crucial in baryon spectroscopy \cite{Glozman:1996wq}. 
For a given meson $\gamma$ the meson exchange potential is
\begin{eqnarray}\label{radialform}
V_\gamma (r) &=&
\frac{g_\gamma^2}{4\pi}\frac{1}{12m_i m_j}
\{\theta(r-r_0)\mu_\gamma^2\frac{e^{-\mu_\gamma r}}{ r} \nonumber\\
&-& \frac {4}{\sqrt {\pi}}
\alpha^3 \exp(-\alpha^2(r-r_0)^2)\}
\end{eqnarray}
In the present calculations we use the
parameters of Ref. \cite{Glozman:1996wq} to which we add the $\mu_{D}$ mass and
the coupling constant $\frac{g_{Dq}^2}{4\pi}$. These are 
$$\frac{g_{\pi q}^2}{4\pi} = \frac{g_{\eta q}^2}{4\pi} =
\frac{g_{Dq}^2}{4\pi}= 0.67,\,\,
\frac{g_{\eta ' q}^2}{4\pi} = 1.206 , $$
$$r_0 = 0.43 \, \mbox{fm}, ~\alpha = 2.91 \, \mbox{fm}^{-1},~~
 C= 0.474 \, \mbox{fm}^{-2}, \, $$
$$ \mu_{\pi} = 139 \, \mbox{ MeV},~ \mu_{\eta} = 547 \,\mbox{ MeV},~
\mu_{\eta'} = 958 \, \mbox{ MeV},\,$$
$$ \mu_{D} = 1867 \, \mbox{ MeV}.  $$
The meson masses correspond to the experimental values from  the  Particle Data Group \cite{Tanabashi:2018oca}.
There is no $K$- and $D_s$-exchange in the $uudc \bar c$ pentaquark and, as discussed 
in the following, we ignore  $\eta_c$-exchange. Therefore
the $K$, $D_s$ and $\eta_c$ masses are not needed in these calculations.

For the quark masses we take the values determined variationally in 
Ref. \cite{Pepin:1996id} 
\begin{equation}\label{quarkmass}
 m_{u,d} = 340 \, \mbox{ MeV},~ m_c = 1350 \, \mbox{ MeV}.
\end{equation}
They were adjusted to reproduce the average mass ${\overline M} = (M + 3 M^*)/4$ = 2008 MeV of
the $D$ mesons. With these values one obtains an average mass  of 2985 MeV for $J/\psi$ and $\eta_c$.

\renewcommand\tabcolsep{0.90cm}
\begin{table*}
\parbox{20cm}{\caption[matrix]{\label{FOURQ}
The hyperfine interaction $ V_{\chi}  $, Eq. (\ref{VCHI}), 
integrated in the flavor-spin space, for the quark subsystem $uudc$
$I$ =1/2, $S$ = 0. \\
${{V}^{q_a q_b}_{\gamma }}$ are defined in Eq.  (\ref{twobody})
where the upper index ${q_a q_b}$ indicates the flavor of the interacting quark pair .  }} 
\begin{tabular}{c|c}
\hline
    State &   $  V_{\chi}  $ \\[0.5ex]
\hline
$|{{\left[{31}\right]}_{O}\
{\left[{22}\right]}_{F}{\left[{22}\right]}_{S}{\left[{4}\right]}_{FS}}\rangle$
  & 15 $ V_{\pi} -  V^{uu}_{\eta} - 2 V^{uu}_{\eta'}  
+ 12 V^{uc}_D - \frac{1}{2} V^{uu}_{\eta_c} + \frac{3}{2} V^{uc}_{\eta_c} - 2  V^{uc}_{\eta'} $ 
\\[1.0ex]
$|{{\left[{31}\right]}_{O}\
{\left[{31}\right]}_{F}{\left[{31}\right]}_{S}{\left[{4}\right]}_{FS}}\rangle$
 & 3 $V_{\pi} + V^{uu}_{\eta} + \frac{1}{3} V^{uu}_{\eta'}
 + 14 V^{uc}_D + \frac{1}{2} V^{uu}_{\eta_c} + \frac{5}{2} V^{uc}_{\eta_c} - \frac{10}{3} V^{uc}_{\eta'}$ 
\\[1.0ex]
$|{{\left[{4}\right]}_{O}\
{\left[{211}\right]}_{F}{\left[{22}\right]}_{S}{\left[{31}\right]}_{FS}}\rangle$
  & 7 $V_{\pi}- \frac{7}{9} V^{uu}_{\eta} - \frac{14}{9} V^{uu}_{\eta'} + 
 + \frac{22}{3}V^{uc}_D - \frac{7}{18} V^{uu}_{\eta_c} + \frac{11}{6} V^{uc}_{\eta_c} - \frac{22}{9} V^{uc}_{\eta'}$
\\[0.9ex]
\hline
\end{tabular}
\end{table*}

After integrating in the flavor space, the two-body matrix elements  
which generalize Eq. (3.3) of Ref. \cite{Glozman:1995fu}  to SU(4), one obtains a result
containing new contributions due to charm. This is 
\begin{equation}\label{twobody}
V_{ij} =
{\bm {\sigma}}_i\ \cdot {\bm {\sigma}}_j\, 
\left\{ \renewcommand{\arraystretch}{2}
\begin{array}{cl}
 V_{\pi} + \frac{1}{3} V^{uu}_{\eta} + \frac{1}{6} V^{uu}_{\eta_c},  &\hspace{0.3cm} \mbox{ $[2]_F, I = 1$} \\
 2 V_K - \frac{2}{3} V^{us}_{\eta}, ~~~ 2 V^{uc}_D- \frac{1}{2} V^{uc}_{\eta_c} & \hspace{0.3cm} \mbox{ $[2]_F, I = \frac{1}{2}$} \\
 2 V^{sc}_{D_s} - \frac{1}{2} V^{sc}_{\eta_c} & \hspace{0.3cm} \mbox{ $[2]_F, I = 0$} \\
 \frac{4}{3} V^{ss}_{\eta} + \frac{3}{2} V^{cc}_{\eta_c} & \hspace{0.3cm} \mbox{ $[2]_F, I = 0$} \\
-2 V^{sc}_{D_s} - \frac{1}{2} V^{sc}_{\eta_c} & \hspace{0.3cm} \mbox{ $[11]_F, I = 0$} \\
-2 V_K - \frac{2}{3} V^{us}_{\eta},~~~ - 2 V^{uc}_D - \frac{1}{2} V^{uc}_{\eta_c} &\hspace{0.3cm} \mbox{ $[11]_F, I = \frac{1}{2}$} \\
-3 V_{\pi} + \frac{1}{3} V^{uu}_{\eta} + \frac{1}{6} V^{uu}_{\eta_c},  &\hspace{0.3cm} \mbox{ $[11]_F, I = 0$}
\end{array} \right.
\end{equation}
\noindent
In Eqs. (\ref{twobody}) the pair  of quarks $ij$ is either in a symmetric [2]$_F$ or in an antisymmetric [11]$_F$ flavor state
and the isospin $I$ is defined by the quark content. The upper index of 
$V$ exhibits the flavor of the
two quarks interchanging a meson specified by the lower index.
In order to keep close to the notations of Ref.  \cite{Glozman:1995fu} the upper index of $\pi$ and 
$K$ is not indicated. Obviously, in every sum/difference of Eq. (\ref{twobody}) the upper index is the same
for all terms.

Note that the term $\frac{1}{2} V^{uc}_{\eta_c}$ was missing in Ref. \cite{Pepin:1996id}. In practice 
it can be neglected but theoretically it is important because it recovers the exact SU(4) limit 
exhibited in Table \ref{exactlimit}.

Using the flavor wave functions given in 
\ref{flavor4} and the expressions (\ref{twobody}) 
one can calculate
the matrix elements of the flavor-spin interaction (\ref{VCHI})
for four quark states. They are presented  in Table \ref{FOURQ}.
To recover the SU(4) limit of Table \ref{exactlimit}
for the $uudc$ configuration, one has to take $V_{\pi}$ = $V^{uu}_{\eta}$ 
= $V^{uu}_{\eta_c}$ = $V^{uc}_D$ = $V^{uc}_{\eta_c}$ = - ${C}_{\chi }$ and $V^{uu}_{\eta'}$ = $V^{uc}_{\eta'}$ = 0.

In the exact SU(4) limit the states $[2]_F$ and $[11]_F$ of Eq. (\ref{twobody}) containing
$u,d,c$ quarks become - 3/2 ${C}_{\chi }$ and 5/2 ${C}_{\chi }$ respectively. Then, making use of the spin matrix elements
 $\langle \sigma_i \cdot \sigma_j\rangle $ = 1  and $\langle \sigma_i \cdot \sigma_j\rangle $ = - 3
for $S = 1$ and  $S = 0$ respectively, one finds that  the contribution of the FS
interaction to the baryon masses are  $V_{\chi}(N)$ =  $V_{\chi}(\Lambda_c$) = - 27/2 $C_{\chi}$
and $V_{\chi}(\Delta)$ = - 9/2 $C_{\chi}$, which are close to the SU(3) limits
- 14 $C_{\chi}$ and - 4 $C_{\chi}$ respectively. As a matter of fact
one can also see that the SU(4) limit given
in - $C_{\chi}$ units, as   
listed in Table \ref{exactlimit}, namely 27, 21 and 15 are quite close to the SU(3) limit
(three flavors) of 28, 64/3 and 16 respectively, which can easily be obtained using 
\ref{Casimir}.

The exact SU(3) or SU(4) limits are useful in indicating the sequence of positive and negative parity states in the
spectrum, which is maintained at a broken unitary symmetry.

\section{Orbital space}\label{coord}

This section follows closely the derivation of the orbital part of the 
four-quark wave function of symmetry $[31]_O$ as presented in Ref. \cite{Stancu:1998sm}.

One needs four internal Jacobi coordinates for pentaquarks. The convenient choice is
\begin{eqnarray}\label{coordin}
\begin{array}{c}\bm{x}\ =\ {\bm{r}}_{1}\ -\ {\bm{r}}_{2}\ ,\, \\ 
\bm{y}\ =\
{\left({{\bm{r}}_{1}\ +\ {\bm{r}}_{2}\ -\ 2{\bm{r}}_{3}}\right)/\sqrt
{3}} ,\, \\
\bm{z}\ =\ {\left({{\bm{r}}_{1}\ +\ {\bm{r}}_{2}\ +\ {\bm{r}}_{3}\ -\
3{\bm{r}}_{4}}\right)/\sqrt {6}} ,\, \\  
\bm{t}\ =\
{\left({{\bm{r}}_{1}\
+\ {\bm{r}}_{2}\ +\ {\bm{r}}_{3}+\ {\bm{r}}_{4}-\
4{\bm{r}}_{5}}\right)/\sqrt {10}} ,
\end{array}
\end{eqnarray}
where 1,2,3 and 4 are the quarks and 5 the antiquark so that
$t$ gives the distance between the antiquark and the center of mass coordinate 
of the four-quark subsystem. Then one has to 
express the orbital wave functions of the four-quark subsystem in terms of the internal coordinates 
$\vec{x},  \vec{y},  \vec{z}$ 
for the specific permutation symmetry
${\left[{31}\right]}_{O}$. For the lowest positive parity states we have considered
an $s^3p$ structure for ${\left[{31}\right]}_{O}$ and extended Moshinski's
method \cite{MO69}, from three  to four particles,
to construct translationally invariant states of definite permutation symmetry.     
In this way we have obtained 
three independent states  denoted below by $\psi_i$,  defined as Young-Yamanouchi
basis vectors \cite{Stancu:1991rc}  of the irreducible representation 
${\left[{31}\right]}_O$   in 
terms of shell model states 
$ \left\langle{\vec{r}\left|{n \ell m}\right.}\right\rangle$ where $n$ = 0, $\ell$ = 1, 
and, for simplicity, we took m = 0 everywhere.
They read
\begin{eqnarray}
{\psi }_{1} = \renewcommand{\arraystretch}{0.5}
\begin{array}{c} $\fbox{1}\fbox{2}\fbox{3}$ \\
$\fbox{4}$\hspace{9mm} \end{array}
=
\left\langle{\vec{x}\left|{000}\right.}\right
\rangle\left\langle{\vec{y}\left|{000}\right.}\right\rangle\left
\langle{\vec{z}\left|{010}\right.}\right\rangle
\end{eqnarray}

\begin{eqnarray}
{\psi }_{2} = \renewcommand{\arraystretch}{0.5}
\begin{array}{c} $\fbox{1}\fbox{2}\fbox{4}$ \\
$\fbox{3}$\hspace{9mm} \end{array}
=
\left\langle{\vec{x}\left|{000}\right.}\right
\rangle\left\langle{\vec{y}\left|{010}\right.}\right\rangle\left
\langle{\vec{z}\left|{000}\right.}\right\rangle
\end{eqnarray}

\begin{eqnarray}
{\psi }_{3} = \renewcommand{\arraystretch}{0.5}
\begin{array}{c} $\fbox{1}\fbox{3}\fbox{4}$ \\
$\fbox{2}$\hspace{9mm} \end{array}
=
\left\langle{\vec{x}\left|{010}\right.}\right
\rangle\left\langle{\vec{y}\left|{000}\right.}\right\rangle\left
\langle{\vec{z}\left|{000}\right.}\right\rangle
\end{eqnarray}

\noindent 
One can see 
that the $\ell$ = 1 orbital excitation is located in the relative motion between quarks
defined by the internal coordinates $\vec{x}$, $ \vec{y}$ and  $\vec{z}$. The states ${\psi }_{1}$ and
${\psi }_{2}$ are  
the generalization, from three to four quarks,
of  $\psi^{\lambda}_{\ell 0}$ and  ${\psi }_{3}$ the generalization, from three to four quarks,
of  $\psi^{\rho}_{\ell 0}$ radial wave functions of mixed
symmetry $[21]$ used in baryon spectroscopy.
Thus, a more appealing notation could be $\psi^{\lambda_1}_{\ell 0}$, $\psi^{\lambda_2}_{\ell 0}$ and 
$\psi^{\rho}_{\ell 0}$ instead of ${\psi }_{i}$ with $i$ = 1,2,3.

In this picture there is no
excitation in the relative motion between the cluster of four quarks and the 
antiquark defined by the coordinate $\vec{t}$. Then the  
pentaquark orbital wave functions $\psi_i^5$ are  obtained by multiplying
each $\psi_i$ from above by the wave function
$\left\langle{\vec{t}\left|{000}\right.}\right\rangle$ which describes the
relative motion between the four-quark subsystem and the antiquark $\overline{c}$.
Assuming an exponential behavior we introduce two variational parameters, 
$a$ for the internal motion of the four-quark subsystem  and
$b$ for the relative motion between the subsystem $qqqc$ and $\overline{c}$. We explicitly have
\begin{equation}\label{psi1}
{\psi }_{1}^{5}\ =\ N\ \exp\ \left[{-\ {\frac{a}{2}}\ \left({{x}^{2}\ +\
{y}^{2}\ +\ {z}^{2}}\right)\ -\ {\frac{b}{2}}\ {t}^{2}}\right]\ z\ {Y}_{10}\
\left({\hat{z}}\right)
\end{equation}
\begin{equation}\label{psi2}
{\psi }_{2}^{5}\ =\ N\ \exp\ \left[{-\ {\frac{a}{2}}\ \left({{x}^{2}\ +\
{y}^{2}\ +\ {z}^{2}}\right)\ -\ {\frac{b}{2}}\ {t}^{2}}\right]\ y\ {Y}_{10}\
\left({\hat{y}}\right)
\end{equation}
\begin{equation}\label{psi3}
{\psi }_{3}^{5}\ =\ N\ \exp\ \left[{-\ {\frac{a}{2}}\ \left({{x}^{2}\ +\
{y}^{2}\ +\ {z}^{2}}\right)\ -\ {\frac{b}{2}}\ {t}^{2}}\right]\ x\ {Y}_{10}\
\left({\hat{x}}\right)
\end{equation}
where
\begin{equation}
N\ =\ {\frac{{2}^{3/2}{a}^{11/4}{b}^{3/4}}{{3}^{1/2}{\pi }^{5/2}}}
\end{equation}

We also need the orbital wave function of 
the lowest negative parity state described by the
$s^4$ configuration of symmetry $[4]_O$  which is
\begin{equation}\label{phi}
\phi_0 = \ N_0\ \exp\ \left[{-\ {\frac{a}{2}}\ \left({{x}^{2}\ +\
{y}^{2}\ +\ {z}^{2}}\right)\ -\ {\frac{b}{2}}\ {t}^{2}}\right],
\end{equation}
with
\begin{equation}\label{phinorm}
N_0\ =\ ({\frac{a}{\pi}})^{9/4} (\frac{b}{\pi})^{3/4}.
\end{equation}

\section{Analytic details}\label{analytic}

The kinetic energy part $T$  of the Hamiltonian (\ref{ham})
can be calculated analytically.  For the positive 
parity states of Table \ref{exactlimit} its expectation value becomes
\begin{eqnarray}
\begin{array}{lcl}\left\langle{T}\right\rangle\ &=&\ {\frac{1}{3}}\
\left[{\left\langle{{\psi }_{1}^{5}\left|{T}\right|{\psi
}_{1}^{5}}\right\rangle\ +\ \left\langle{{\psi }_{2}^{5}\left|{T}\right|{\psi
}_{2}^{5}}\right\rangle\ +\ \left\langle{{\psi }_{3}^{5}\left|{T}\right|{\psi
}_{3}^{5}}\right\rangle}\right]\\
\\ &=&\ {\hbar }^{2}\ \left({{\frac{11}{2{\mu }_{1}}}\ a\ +\ {\frac{3}{{2\mu
}_{2}}}\ b}\right),
\end{array}
\end{eqnarray}
with
\begin{eqnarray}
{\frac{4}{{\mu }_{1}}}\
 =  \ \frac{3}{{m}_{q}}\  +  \ \frac{1}{{m}_{Q}},\ 
\end{eqnarray}
and  
\begin{equation}\label{redmass}
{\frac{5}{{\mu }_{2}}}\ =\ {\frac{1}{{\mu }_{1}}}\ +\ {\frac{4}{{m}_{Q}}},
\end{equation}
where $q = u, d$ and $Q = c$.
Here, we have  $ m_q $ = 340 MeV and $m_c$ = 1350 MeV, 
as defined by Eq. (\ref{quarkmass}). 
Taking $m_u$ = $m_d$  = $m_Q$ = $m$ and setting a = b, one can recover the identical particle limit
$\left\langle{T}\right\rangle\ =\ {\frac{7}{2}}\ \hbar \omega$ with
$\hbar \omega  \ =\ 2\ a{\hbar }^{2}/m$, see \cite{HO}.
Note that one can obtain  the desired units of all quantities involved, by muliplying/dividing  by
$\hbar c \simeq $  197.33 MeV fm at the correct power.

By integrating in the color space, the expectation value of the confinement interaction (\ref{conf}) 
becomes 
\begin{equation}\label{confin}
\left\langle{{V}_{conf}}\right\rangle\ =\ {\frac{C}{2}}\ \left({6\
\left\langle{{r}_{12}}\right\rangle\ +\ 4\
\left\langle{{r}_{45}}\right\rangle}\right)
\end{equation}
where the coefficients 6 and 4 account for the number of quark-quark and
quark-antiquark pairs, respectively, and
\begin{equation}\label{r12}
\left\langle{{r}_{ij}}\right\rangle\ =\ {\frac{1}{3}}\
\left[{\left\langle{{\psi }_{1}^{5}\left|{{r}_{ij}}\right|{\psi
}_{1}^{5}}\right\rangle\ +\ \left\langle{{\psi
}_{2}^{5}\left|{{r}_{ij}}\right|{\psi }_{2}^{5}}\right\rangle\ +\
\left\langle{{\psi }_{3}^{5}\left|{{r}_{ij}}\right|{\psi
}_{3}^{5}}\right\rangle}\right],
\end{equation}
where $i,j$ = 1,2,3,4,5 ($ i \neq j$).
An analytic evaluation gives
\begin{equation}
\left\langle{{r}_{12}}\right\rangle\ =\ {\frac{20}{9}}\ \sqrt {{\frac{1}{\pi
 a}}},
\end{equation}
as well as for ${r}_{ij}$ with  $i,j$ = 1,2,3,4 and 
\begin{equation}
\left\langle{{r}_{45}}\right\rangle\ =\ {\frac{1}{3\sqrt {2\pi }}}\
\left[{2\sqrt {{\frac{3}{a}}\ +\ {\frac{5}{b}}}\ +\ \sqrt {5b}\
\left({{\frac{1}{2a}}\ +\ {\frac{1}{b}}}\right)}\right],
\end{equation}
and likewise for ${r}_{j5}$ where $j$ = 1,2,3.

One also needs to derive the expressions of $\left\langle{{r}_{12}}\right\rangle$
and $\left\langle{{r}_{45}}\right\rangle$  in the case where the four quarks are in the
$s^4$ configuration. These are
\begin{equation}\label{r12ground}
\left\langle{{r}_{12}}\right\rangle\ = \sqrt {{\frac{4}{\pi a}}},
\end{equation}
and 
\begin{equation}\label{r45ground}
\left\langle{{r}_{45}}\right\rangle\ =\ {\frac{1}{\sqrt {2\pi }}}\
\sqrt{{\frac{3}{a}}\ +\ {\frac{5}{b}}}.
\end{equation}

The matrix elements of the spin-flavor operators of Eq. (\ref{VCHI}) have been calculated
using the fractional parentage technique described in Ref. \cite{Stancu:1991rc} based on
Clebsch-Gordan coefficients of the group $S_4$ \cite{Stancu:1999qr}. In this way, each
matrix element reduces to a linear combination of two-body matrix elements of
either symmetric or antisymmetric states for which 
Eq. (\ref{twobody})
can be used to integrate in the spin-flavor space. 
For positive parity states with one unit of orbital excitation the resulting
linear combinations contain orbital two-body matrix elements of type
$\left\langle{ss\left|{{V}^{q_a q_b}_{\gamma }}\right|ss}\right\rangle\ ,\
\left\langle{sp\left|{{V}^{q_a q_b}_{\gamma }}\right|sp}\right\rangle$ and
$\left\langle{sp\left|{{V}^{q_a q_b}_{\gamma }}\right|ps}\right\rangle$ where $\gamma =
\pi, D, \eta, \eta '$  and $q_a q_b$
is a pair of quarks from Eq. (\ref{twobody}).

It is useful to note  that the ratio of the orbital  matrix 
elements of the $D$ meson exchange interaction between the pair $uc$ of quarks, denoted here 
by $V^{uc}_D$ and the matrix elements of the $\pi$ meson exchange interaction $V_{\pi}$, see Eq. (\ref{twobody}), 
turned out to be about 0.25, in agreement with the mass ratio $m_{u,d}/m_c$ of Eq. (\ref{quarkmass}).


\section{Results}\label{numerical}

The four quark states described in the previous sections couple
to the antiquark to a total angular momentum  $\vec{J}\ =\ \vec{L}\ +\ \vec{S}\ +\
{\vec{s}}_{Q}$, with $\vec{L}$, $\vec{S}$ the angular momentum and spin of the
four-quark cluster and $\vec{s}_Q$ the spin of the heavy antiquark.

We have searched for  variational solutions for the pentaquark
described by the Hamiltonian defined in Sec. \ref{Hamiltonian}.
The numerical results are presented in Table  \ref{results}.
The wave functions are  the product of the four quarks subsystem states 
of flavor-spin structure defined in Table \ref{FOURQ}  and the
charm antiquark wave function denoted by $|\bar c \rangle$.  The orbital  wave 
functions of the pentaquark are given by Eqs. (\ref{psi1})-(\ref{phinorm})
containing the variational parametres $a$ and $b$. 
We have neglected the contribution of  $V^{uu}_{\eta_c}$ and of $V^{uc}_{\eta_c}$, because little
$u \bar u$ and $d \bar d$ is expected in ${\eta_c}$.   We have also neglected 
$V^{uc}_{\eta'}$ assuming a little $c \bar c$ component in $\eta'$, which means that 
we took 
\begin{equation}
V^{uu}_{\eta_c} = V^{uc}_{\eta_c} = V^{uc}_{\eta'} = 0.
\end{equation}

\renewcommand\tabcolsep{0.60cm}
\begin{table*}
\caption{\label{results}Lowest positive and negative parity  $uudc \bar c$ pentaquarks
of quantum numbers $I,~ J^P$ and symmetry structure defined in Table \ref{exactlimit}.
Column 1 gives the state, column 2 the isospin, column 3  the parity and total angular momentum, 
column 4 the
optimal variational parameters associated to the wave functions defined in Sec. \ref{coord},
column 5 the calculated mass 
and the last two columns the distances between quarks/antiquarks.}
\begin{tabular}{cccccccc}
\hline
State & $I$ & $J^P$ & \multicolumn{2}{c}{Variational parameters} &
Mass  & $\langle r_{12} \rangle$ & $\langle r_{45} \rangle$\\
& & \hspace{6mm} & a (fm$^{-2}$) & b (fm$^{-2}$) & (GeV) & fm & fm \\
\hline
$ | 1 \rangle ~|\overline c \rangle$ &  $\frac{1}{2}$ &   $\frac{1}{2}^+$, $\frac{3}{2}^+$ 
&  2.055 & 1.079 & 4.273 &   0.875 & 1.018 \\[1.1ex]
$ | 2 \rangle ~|\overline c\rangle$  &  $\frac{1}{2}$,~$\frac{3}{2}$ &   $\frac{1}{2}^+$,\hspace{1mm} $\frac{3}{2}^+$,\hspace{1mm} $\frac{5}{2}^+$ 
&  1.541 & 1.027 & 4.453 & 1.010 & 1.085 \\[1.1ex]
$ | 3 \rangle ~|\overline c \rangle$ &  $\frac{1}{2}$ &   $\frac{1}{2}^-$ 
&  1.284 & 1.079 & 4.487 &   0.996 & 1.062 \\[1.1ex]
\hline
\end{tabular}
\end{table*}

The optimal values found for 
the parameters $a$ and $b$ are indicated in each case. The eigenvalues of  $|1 \rangle |\bar c \rangle $
and    $|2 \rangle |\bar c \rangle $  states are degenerate for the allowed values of $J$ in each case.  
Moreover, for the wave function   $|2 \rangle |\bar c \rangle $ the states with $J^P =   \frac{1}{2}^+$ and  $\frac{3}{2}^+$
have multiplicity 2. 
Presently there are more states than the observed ones.
In the case of pentaquarks with heavy flavor, Ref. \cite{Ali:2019npk} argues that only part of the spectrum 
is reachable in $\Lambda_b$ decays.

The SU(4) FS model described in Sec. \ref{Hamiltonian} 
gives a value for the  mass of the state $|1 \rangle |\bar c \rangle $ of 4273 MeV, close to the mass of the
$P^+_c(4312)$ resonance and supports the assignment $J^P =   \frac{1}{2}^+$ or  $\frac{3}{2}^+$ for it.
The state $|2 \rangle |\bar c \rangle $ has a mass of 4453 MeV close to that of the 
$P^+_c(4440) $ and  $P^+_c(4457) $  resonances  and supports the quantum numbers  $J^P = \frac{1}{2}^+, \frac{3}{2}^+$ or  $\frac{5}{2}^+$.

As discussed above, 
in the exact limit one obtains binding for $|1 \rangle |\bar c \rangle $, see Eq. (\ref{state1}),
but this is not the
case for the variational solution in the broken SU(4), which is larger than the 
lowest thresholds, $N\eta_c(3920)$ or $NJ/\psi(4040)$.
It is not surprising, because in the exact limit the matrix elements are equal for
all exchanged mesons, which means that the matrix elements of $\eta$, $\eta'$ and  $D$
exchanges are overestimated, thus introducing more attraction. However 
the SU(4) exact limit brings useful information about the relative position
of various states. 

The lowest negative parity state $|3 \rangle |\bar c \rangle $
is located above the positive parity states, as expected. It has a mass of 4487 MeV close 
that of the $P^+_c(4440)$ or the  $P^+_c(4457) $ resonances. It supports the quantum number $J^P$ = $1/2^-$.

It may be useful to look at the compactness of the present pentaquarks by estimating 
the distances $\left\langle{{r}_{12}}\right\rangle$ and $\left\langle{{r}_{45}}\right\rangle$ 
from Eqs. (\ref{r12})-(\ref{r45ground}). They are exhibited in Table \ref{results}.
In the $|1 \rangle |\overline c \rangle $ state, which is the lowest, one can see that at equilibrium the four quarks 
are clustered together and the antiquark is somewhat far apart. For the other states the 
distances  $\left\langle{{r}_{12}}\right\rangle$  and $\left\langle{{r}_{45}}\right\rangle$ 
are comparable.

The values of the above convenient relative distances were exhibited in order to see whether or not some clustering of 
four-quark formation is possible. The effect is quite small.  
One can use alternative relative coordinates convenient either for a molecular-type structure,
or for the hadrocharmonium picture. They are both 
deuteron-like. Unitary transformations can relate various coordinate systems.  This can be considered in the future. 

Also, one could introduce a third variational 
parameter, different from $a$ or $b$, associated to the coordinate $z$ where the charm quark is involved. 
In that case the analytic work would be more involved. It was found that the energy levels vary slowly 
with the present variational parameters, so that the minima are very shallow. In view of this observation  
the spectrum is expected to be a little 
lowered by the introduction of a new parameter, but the parity sequence will remain the same.
However a more sophisticated variational solution would be important for the wave function,
thus for studying decay properties numerically. More theoretical work should be done, in wait for
more precise data.

\section{Strong decay widths}\label{width}

Assuming a coupling to open channels, which could provide estimates of widths (and level shifts),
one can qualitatively discuss the strong decay widths in the context of compact 
pentaquarks. First, for P-waves the centrifugal barrier  will reduce the widths
relative to S-waves. Thus, in the 
present model, the lowest state, having positive parity, will have a smaller width than the 
negative parity ones, contrary to the arguments of Ref. \cite{Burns:2015dwa}
inspired by OGE or molecular models. Next, the narrowness of the lowest state is related to the 
overlap of the color-flavor-spin-orbital wave function with an $p + J/\psi$ final state. 
For the lowest state, 
$|{{\left[{31}\right]}_{O}\
{\left[{22}\right]}_{F}{\left[{22}\right]}_{S}{\left[{4}\right]}_{FS}}\rangle$,
the overlap with $q^3$ + $Q \bar Q$ may be small. This problem deserves a separate investigation.


\section{Comparison with other studies}\label{comp}

The basis (\ref{psi1}) - (\ref{psi3}) is entirely consistent with the parity 
considerations made
in the context of the tetrahedral group $T_d$  \cite{Bijker:2003pm},
isomorphic with the permutation group $S_4$. 
The states $|1 \rangle$ and $|2 \rangle$ of Table \ref{exactlimit}
are states of symmetry $1^-_{A_1}$ 
in the notation of Ref. \cite{Bijker:2003pm}. Likewise the  state $|3 \rangle$
of the same table has symmetry $0^+_{A_1}$. 
There is agreement between 
our findings and the  conclusion of  Ref. \cite{Bijker:2003pm} that the parity 
of a pentaquark depends on the interplay between the flavor-spin interaction 
and the contribution of the orbital excitation to the mass. Here we found  
that the flavor-spin attraction compensates the excess of kinetic energy brought
in by one unit of orbital excitation in the internal motion of the four-quark subsystem containing a charmed quark,
as it was the case for pentaquarks containing only light flavors like $uudd$  \cite{Stancu:1998sm}, so that
the lowest state has positive parity in both cases, either for light or light+heavy quarks.

The authors of Ref. \cite{Yuan:2012wz} have studied the $qqqc \bar c$ pentaquark
prior to the 2015 LHCb observation of pentaquarks, in three different
models, including  the flavor-spin (FS) model. 
The orbital wave function of the four-quark subsystem has symmetry $[4]_O$ for both 
parities. 
A symmetric state with an $s^3p$ structure describes the spurious
centre of mass motion, while the state of orbital symmetry $[31]_O$ derived in Sec. \ref{coord} is 
translationally invariant. 
Although the radial wave function was not specified, one can infer that the positive parity states of Ref.  \cite{Yuan:2012wz} 
were obtained by including a unit of orbital angular momentum in the relative motion  
between the four-quark subsystem and the antiquark.
%
%
In this case the orbital wave function takes the form
\begin{equation}\label{psi4}
{\psi }_{4}^{5}\ =\ N_4 \exp\ \left[{-\ {\frac{a}{2}}\ \left({{x}^{2}\ +\
{y}^{2}\ +\ {z}^{2}}\right)\ -\ {\frac{b}{2}}\ {t}^{2}}\right]\ t\ {Y}_{10}\
\left({\hat{t}}\right),
\end{equation}
where
\begin{equation}
N_4 =\ \frac{8^{1/2} a^{9/4} b^{5/4}}{3^{1/2} \pi^{5/2}}. 
\end{equation}
Then the expectation value of the kinetic energy becomes
\begin{eqnarray}
\begin{array}{lcl}\left\langle{T_t}\right\rangle\ 
 &=&\ {\hbar }^{2}\ \left({{\frac{3}{{m}_{q}}}\ a\ +\ {\frac{3}{8 {\mu_1}}}\ a
+\ {\frac{1}{8 {\mu_2}}}}\ b\right)\end{array},
\end{eqnarray}
with 
\begin{eqnarray}
{\frac{1}{{\mu }_{1}}}\
 =  \ \frac{1}{{m}_{q}}\  +  \ \frac{3}{{m}_{Q}},\ 
\end{eqnarray}
and  
\begin{equation}
{\frac{1}{{\mu }_{2}}}\ =\ {\frac{3}{{m_q }}}\ +\ {\frac{17}{{m}_{Q}}}.
\end{equation}
The expectation value of the confinement interaction is given by Eq. (\ref{confin})
with
\begin{equation}
\left\langle{{r}_{12}}\right\rangle\ =\ \sqrt {{\frac{4}{\pi
 a}}},
\end{equation}
and
\begin{equation}
\left\langle{{r}_{45}}\right\rangle\ =\ {\frac{2}{3} \sqrt{\frac{2\ b}{5\ \pi }}}\ 
\left({{\frac{3}{4a}}\ +\ {\frac{5}{b}}}\right).
\end{equation}
The hyperfine interaction integrated in the flavor-spin space is given
by the third row of Table \ref{FOURQ}
because the symmetry state corresponds to  
$|{{\left[{4}\right]}_{O}\
{\left[{211}\right]}_{F}{\left[{22}\right]}_{S}{\left[{31}\right]}_{FS}}\rangle$.


The attraction brought by the flavor-spin interaction  in 
the four-quark state $[31]_{FS}$ is not strong enough to compensate the excess of kinetic energy due to the orbital
excitation of the antiquark relative to the four quark subsystem. In our case we obtain a mass of 
4573 MeV ($a$ = 0.040 GeV$^2$, $b$ = 0.037 GeV$^2$) which corresponds to $J^P$ = 1/2$^+$ or 3/2$^+$, higher than the
mass 4487 GeV of the 1/2$^-$ state of Table \ref{results}.
This can explain 
why the lowest negative parity state $1/2^-$ in the calculations of Ref. \cite{Yuan:2012wz}
is lower than
the  lowest $1/2^+$ state. 

In Ref. \cite{Yang:2015bmv} the spectrum of the $uudc \bar c$ is studied in a
chiral quark model where the Hamiltonian  contains both the chromomagnetic  and the flavor-spin interactions.
For the flavor-spin part the symmetry is restricted to SU(3) with the replacement of
the $s$ quark by the $c$ quark which, according to Eq. (\ref{twobody}),
seems to be  reasonable. However, the lowest states have negative parity and the 
results for positive parity states are not shown, claiming that they are unbound. 
A possible interpretation is that the chromomagnetic dominates over the flavor-spin interaction in 
those calculations.

However, some  support in favor of a flavor-spin interaction in exotics is given by Ref. \cite{Giron:2019cfc}.
Although restricted to tetraquarks described in a diquark basis, it has been shown that the mass difference
observed in some charmed tetraquarks can be explained by an isospin dependent interaction, which is an
SU(2) flavor-spin interaction. An important remark  is that the contribution of the flavor-spin part relative to the 
one gluon exchange part must be adequate, irrespective of the basis used in the calculations. 

\section{Conclusions}

We have calculated the lowest part of the mass spectrum of the hidden charm 
pentaquark $uudc \bar c$,  within the SU(4) version of the flavor-spin model
introduced in this work. This is a model which provides an internal structure 
for pentaquarks, contrary to the models mentioned in the introduction.
An important achievement is that
for positive parity states
we have made a clear distinction between the case where  a unit of angular 
momentum is located in the internal motion of the four-quark subsystem and the previously schematic case
used in Ref. \cite{Yuan:2012wz}, 
where the angular momentum has to be located in the relative motion between the four-quark subsystem 
and the antiquark in order to have translationally invariant states (no center of mass motion)
of specific flavor-spin symmetry.
We have presented analytic expressions for the kinetic, confinement and
the flavor-spin parts of the  Hamiltonian for a simple variational solution. 
In the numerical estimates we have used model parameters from earlier studies of pentaquarks.

The calculated  masses are in the  range of  the presently observed 
$P^+_c(4312)$,  $P^+_c(4440)$ and   $P^+_c(4440)$ resonances of the  2019 LHCb data
both for positive and negative parity states.

One should stress that an important feature of the flavor-spin model is that it introduces an
isospin dependence of the pentaquark states, necessary to discriminate between decay 
channels, like for tetraquarks \cite{Giron:2019cfc}.

The radial form (\ref{radialform}) favors the $p$ states over 
the $s$ states due to the second term, which makes the FS interaction closer to the SU(4) limit.
Therefore it  would 
be  interesting to find out how the pentaquark spectrum depends on
the particular radial form of the FS hyperfine interaction by choosing 
the more realistic version of Ref. \cite{Glozman:1997jy} 
with a cut-off parameter for each individual meson exchange. 
Based on symmetry arguments we however expect a similar conclusion for other 
forms of the hyperfine interaction.

In addition, one should look for more suitable
coupling constants related to the presence of the charmed quark.

Presently, the most important conclusion is that the FS model predicts positive
parity for the lowest state in contrast to the widely used OGE model,
with or without correlated quarks/antiquarks. 

Among the  recent studies interpreting the 2019 LHCb data, positive 
parity states have been considered in Ref.  \cite{Ali:2019npk}
within a doubly-heavy triquark --- light diquark model. In that model, 
besides the spin-spin chromomagnetic interaction,
the Hamiltonian incorporates spin-orbit and orbital interactions. 
The fitted parametres lead to a positive contribution of the angular momentum
dependent terms which make the $\ell = 1$ states to be higher than the $\ell = 0$ states. 
This means that the lowest state has a negative parity,
contrary to our results.

 
The 2019 LHCb pentaquarks have also been studied as hadrocharmonium states \cite{Eides:2019tgv}.
In that model the new $P^+_c(4312)$ resonance is interpreted as a bound state of $\chi_0$(1P) and 
a nucleon with $I = 1/2$ and $J^P = 1/2^+$. The other two resonances $P^+_c(4440)$ and   $P^+_c(4457)$,
have a negative parity. The parity sequence in the spectrum is  closer to the one 
produced by the flavor-spin model. 

The main issue of the present work is the parity sequence in the pentaquark $ u u d c \bar c$ spectrum. 
Thus, a crucial test is to   experimentally determine the spin and parity of the 
2019 LHCb resonances in order to discriminate  between different interpretations,  in
particular between the OGE and the flavor-spin quark models, the doubly-heavy triquark --- light diquark model,
the molecular picture or  
the hadronic molecular picture. 

\appendix

\section{The SU(4) generators}\label{lambda}

Here we reproduce the $\lambda^F$ matrices which are the 
SU(4) generators in the fundamental representation of SU(4) \cite{Stancu:1991rc}.
Implementing them in Eq. (\ref{VCHI}) one can obtain the two-body matrix elements
of Eq. (\ref{twobody}) for each pair of quarks of a given flavor.

\begin{eqnarray}
\lambda^1   &  = &  
\left( 
\begin{array}{cccc}
0 & 1 & 0 & 0 \\
1 & 0 & 0 & 0 \\ 
0 & 0 & 0 & 0  \\
0 & 0 & 0 & 0 \end{array} 
\right)
,\,\,\,
\lambda^2 = 
\left(
\begin{array}{cccc}
0 & - i & 0 & 0 \\
i & 0 & 0 & 0 \\ 
0 & 0 & 0 & 0  \\
0 & 0 & 0 & 0 \end{array}
\right)
\end{eqnarray} 
\begin{eqnarray}
\lambda^3  &  = & 
\left(
\begin{array}{cccc}
1 & 0 & 0 & 0 \\
0 & - 1 & 0 & 0 \\ 
0 & 0 & 0 & 0  \\
0 & 0 & 0 & 0 \end{array}
\right)
,\,\,\,
\lambda^4 = 
\left(
\begin{array}{cccc}
0 & 0 & 1 & 0 \\
0 & 0 & 0 & 0 \\ 
1 & 0 & 0 & 0  \\
0 & 0 & 0 & 0 \end{array}
\right)
\end{eqnarray}
\begin{eqnarray}
\lambda^5  &  = & 
\left(
\begin{array}{cccc}
0 & 0 & - i & 0 \\
0 & 0 & 0 & 0 \\ 
i & 0 & 0 & 0  \\
0 & 0 & 0 & 0 \end{array}
\right) 
,\,\,\,
\lambda^6 = 
\left(
\begin{array}{cccc}
0 & 0 & 0 & 0 \\
0 & 0 & 1 & 0 \\ 
0 & 1 & 0 & 0  \\
0 & 0 & 0 & 0 \end{array}
\right) 
\end{eqnarray}
\begin{eqnarray}
\lambda^7  &  = & 
\left(
\begin{array}{cccc}
0 & 0 & 0 & 0 \\
0 & 0 & - i & 0 \\ 
0 & i & 0 & 0  \\
0 & 0 & 0 & 0 \end{array}
\right) 
,\,\,\,
\lambda^8    =
\frac{1}{\sqrt{3}}
\left(
\begin{array}{cccc}
1 & 0 & 0 & 0 \\
0 & 1 & 0 & 0 \\ 
0 & 0 & - 2 & 0  \\
0 & 0 & 0 & 0 \end{array}
\right) 
\end{eqnarray}
\begin{eqnarray}
\lambda^9  &  = & 
\left(
\begin{array}{cccc}
0 & 0 & 0 & 1 \\
0 & 0 & 0 & 0 \\ 
0 & 0 & 0 & 0  \\
1 & 0 & 0 & 0 \end{array}
\right) 
,\,\,\,
\lambda^{10}    = 
\left(
\begin{array}{cccc}
0 & 0 & 0 & - i \\
0 & 0 & 0 & 0 \\ 
0 & 0 & 0 & 0  \\
i & 0 & 0 & 0 \end{array}
\right) 
\end{eqnarray}
\begin{eqnarray}
\lambda^{11}  &  = & 
\left(
\begin{array}{cccc}
0 & 0 & 0 & 0 \\
0 & 0 & 0 & 1 \\ 
0 & 0 & 0 & 0  \\
0 & 1 & 0 & 0 \end{array}
\right) 
,\,\,\,
\lambda^{12} = 
\left(
\begin{array}{cccc}
0 & 0 & 0 & 0 \\
0 & 0 & 0 & - i \\ 
0 & 0 & 0 & 0  \\
0 & i & 0 & 0 \end{array}
\right) 
\end{eqnarray}
\begin{eqnarray}
\lambda^{13}  &  = & 
\left(
\begin{array}{cccc}
0 & 0 & 0 & 0 \\
0 & 0 & 0 & 0 \\ 
0 & 0 & 0 & 1  \\
0 & 0 & 1 & 0 \end{array}
\right) 
,\,\,\,
\lambda^{14} = 
\left(
\begin{array}{cccc}
0 & 0 & 0 & 0 \\
0 & 0 & 0 & 0 \\ 
0 & 0 & 0 & - i  \\
0 & 0 & i & 0 \end{array}
\right) 
\end{eqnarray}
\begin{eqnarray}
\lambda^{15}  &  = & 
\frac{1}{\sqrt{6}}
\left(
\begin{array}{cccc}
1 & 0 & 0 & 0 \\
0 & 1 & 0 & 0 \\ 
0 & 0 & 1 & 0  \\
0 & 0 & 0 & - 3 \end{array}
\right) 
\end{eqnarray}

\section{Exact SU(4) limit}\label{Casimir}

The expectation value of the operator (\ref{schematic}) 
displayed in Table \ref{exactlimit}, can be checked with the following
formula \cite{Ortiz-Pacheco:2018ccl}
\begin{eqnarray}
\langle ~\sum_{i<j} \lambda^F_i \cdot \lambda^F_j \bm{\sigma}_i \cdot \bm{\sigma}_j~\rangle 
&=& 4 C_{2}^{SU(2n)} - 2 C_{2}^{SU(n)} - \frac{4}{k} C_{2}^{SU(2)} \nonumber \\ 
&-& k \frac{3(n^2 - 1)}{n}
\label{lambdasigma}
\end{eqnarray}
where $n$ is the number flavors and  $k$ the number of quarks, here $n$ = 4 and $k$ = 4.  $C_{2}^{SU(n)}$ is the
Casimir operator eigenvalues of $SU(n)$ which can be derived from the
expression  \cite{Stancu:1997dq} :
\begin{eqnarray}\label{casimiroperator}
C_{2}^{SU(n)} &=& \frac{1}{2} [f_1'(f_1'+n-1) + f_2'(f_2'+n-3) + f_3'(f_3'+n-5)
 \nonumber \\
&+& f_4'(f_4'+n-7) + ... + f_{n-1}'(f_{n-1}'-n+3) ] \nonumber \\
&-& \frac{1}{2n} (\sum_{i=1}^{n-1} f_i')^2
\label{casimir}
\end{eqnarray}
where $f_i'= f_i-f_n$, for an irreducible representation given by the
partition $[f_1,f_2,...,f_n]$.
Eq. (\ref{lambdasigma}) has been previously used for $n$ = 3 and $k$ = 6 in Ref.  \cite{Stancu:1997dq}. 

\section{The baryons}\label{baryons}

The masses of baryons related to the present study were calculated variationally 
using a radial wave function of the form 
$\phi \propto  exp[-\frac{a}{2}(x^2 + y^2)]$
containing the variational parameter  $a$
and the coordinates $x$ and $y$ defined by Eq. (\ref{coordin}).

The results are indicated in Table \ref{baryon} together with the experimental
masses. One can see that condition  
$V^{uc}_{\eta'} \neq 0$ is disfavored because in this case the mass 
difference between $\Sigma_c^*$ and  $\Sigma_c$ is negligible as compared to the experiment.
This quantitatively 
justify the condition $V^{uc}_{\eta'} = 0$, imposed in the case of pentaquarks.
(Note that the quantity $V^{uc}_{\eta'}$ is not present in the mass of $\Lambda_c$.)

The calculated charmed baryon masses with  $V^{uc}_{\eta'} = 0$   are 
somewhat smaller than the experimental values. By increasing the charmed 
quark mass from $m_c$ = 1.35 GeV to $m_c$ = 1.45 GeV the agreement
with the experiment is much better. If the same increase is
taken into account in calculating the pentaquark masses, the spectrum 
is shifted to larger masses but still remains at the edge of the desired experimental
range. However the conclusion remains unchanged: the lowest state
has positive parity.

\renewcommand\tabcolsep{0.60cm}
\begin{table*}
\caption{\label{baryon} Variational solution for the baryon ground state masses with the
flavor-spin interaction of Sec. \ref{Hamiltonian}.
Column 1 gives the baryon, column 2 the isospin, column 3 the spin and parity 
column 4 the mass with $V^{uc}_{\eta'} = 0$ and $V^{uc}_{\eta'} \neq 0$, column 5 the variational parameter 
and the last column the experimental mass.}
\begin{tabular}{ccccccc}
\hline
Baryon & \hspace{2mm}$I$ & $J^P$ & \multicolumn{2}{c}{Calculated Mass (GeV)}     & \hspace{1mm} a(fm$^{-2}$) & \hspace{0mm} Exp.mass (GeV)\\
&\hspace{6mm} & \hspace{6mm} & \hspace{2mm} $V^{uc}_{\eta'} = 0$ & $V^{uc}_{\eta'} \neq 0$   &  & \\
\hline
$ N $ & \hspace{1mm} $\frac{1}{2}$ & \hspace{2mm}  $\frac{1}{2}^+$ 
& \hspace{5mm} 0.960 &  & \hspace{2mm} 2.594  & 0.940 \\[1.1ex]
$ \Delta $  & \hspace{1mm} $\frac{3}{2}$ & \hspace{2mm} $\frac{3}{2}^+$ & \hspace{5mm} 1.304 &  & \hspace{2mm} 2.594  & 1.232 \\[1.1ex]
$ \Lambda_c $ & \hspace{1mm} $ 0 $ & \hspace{2mm}  $\frac{1}{2}^+$ & \hspace{5mm} 2.180 &  & \hspace{2mm}  2.055 
&   2.283  \\[1.1ex]
$ \Sigma_c $ & \hspace{1mm} $ 1 $ & \hspace{2mm}  $\frac{1}{2}^+$ & \hspace{5mm} 2.346 & \hspace{2mm}2.434 & \hspace{2mm} 2.055
&   2.455  \\[1.1ex]
$ \Sigma_c^* $ & \hspace{1mm} $ 1 $ & \hspace{2mm}  $\frac{3}{2}^+$ & \hspace{5mm} 2.482 & \hspace{2mm}2.438 & \hspace{2mm} 2.055
&   2.530  \\[1.1ex]
\hline
\end{tabular}
\end{table*}

In addition, we have calculated the masses of two  doubly charmed baryons, 
namely $\Xi^{++}_{cc}({\frac{1}{2}})$ and $\Xi^{++}_{cc}({\frac{3}{2}})$. They are 3386 MeV and 3520 MeV respectively,
when $V^{uc}_{\eta'}$ = 0. From experimental point of view the situation remains controversial.
Recently the LHCb collaboration announced the obsevation of the 
double charm baryon $\Xi^{++}_{cc}$ in the $\Lambda^+_c K^- \pi^+ \pi^+$ mass spectrum
\cite{Aaij:2017ueg}.
Its mass of 3621 MeV is at variance with the 3519 MeV value found earlier by  the SELEX 
Collaboration  \cite{Mattson:2002vu}. The spin and parity are not known experimentally.


\section{The flavor wave functions}\label{flavor4}

Here we give the explicit form of flavor states 
of content $uudc$ in the Young Yamanouchi basis of irreducible representations (irrep) $[{f}]_F$
appearing in the wave functions. 
We assume identical particle and use the method of Ref. \cite{Stancu:1991rc}
to derive these basis vectors. The order of particles is always 1234 in every term below
and the permutation symmetry of each state is defined by its Yamanouchi
symbol, which is a compact notation for a Young tableau. For a tableau with $n$ particles it
is defined by $Y = (r_n,r_{n-1},...,r_1)$ where $r_i$ represents the row of the particle $i$.

For the irrep $[22]$  there are two basis vectors
\begin{eqnarray}\label{22s}
| [22]_F 2211 \rangle  &=& \sqrt{\frac{1}{6}} (uudc + uucd + dcuu + cduu \nonumber \\
& - & \frac{1}{2}duuc - \frac{1}{2}cuud - \frac{1}{2}ucdu - \frac{1}{2}udcu -  \frac{1}{2}uduc  \nonumber \\
& - & \frac{1}{2}ducu - \frac{1}{2}ucud -  \frac{1}{2} cudu)
\end{eqnarray}
\begin{eqnarray}\label{22a}
| [22]_F 2121 \rangle  &=& \sqrt{\frac{1}{8}}[ (ud - du)(uc - cu)\nonumber \\
& + & (uc - cu)(ud - du)]
\end{eqnarray}

For the irrep $[31]$  there are three basis vectors
\begin{eqnarray}\label{31s}  
| [31]_F 2111 \rangle  &=&  \frac{1}{6} (3uudc + 3duuc + 3 uduc  \nonumber \\
& - & cudu - cuud  - cduu - ucdu - dcuu \nonumber \\
& - & ucud - uucd - ducu - udcu),
\end{eqnarray}
\begin{eqnarray}\label{31m}
| [31]_F 1211 \rangle  &=& \sqrt{\frac{1}{18}} (2uucd + 2ducu + 2udcu  \nonumber \\
& - & cuud - cudu - cduu \nonumber \\
& - & ucud - dcuu - ucdu),
\end{eqnarray}
\begin{eqnarray}\label{31a} 
| [31]_F 1121 \rangle  &=& \sqrt{\frac{1}{6}} (ucud + dcuu + ucdu \nonumber \\
& - & cuud - cduu - cudu)
\end{eqnarray}
Note that there is a conflict with the basis vectors (A.9)-(A.11) of Ref. \cite{Yuan:2012wz}
which do not form a Young Yamanouchi basis, thus, although orthogonal, they do not satisfy the correct 
permutation symmetry. 

In the same way the three basis vectors of the irrep $[211]$ were obtained as
\begin{eqnarray}\label{211s}  
| [211]_F 3211 \rangle  &=&  \frac{1}{4} (2uudc - 2uucd - uduc - duuc + udcu \nonumber \\
& + & ducu + ucud + cuud - ucdu - cudu),
\end{eqnarray}
\begin{eqnarray}\label{211a} 
| [211]_F 1321 \rangle  &=& \sqrt{\frac{1}{6}} (udcu + cudu + dcuu \nonumber \\
& - & cduu - ucdu - ducu),
\end{eqnarray}
\begin{eqnarray}\label{211m}
| [211]_F 3121 \rangle  &=& \sqrt{\frac{1}{48}} (3uduc - 3duuc + 3cuud - 3ucud \nonumber \\
& + &  2dcuu - 2cduu - cudu + ucdu \nonumber \\
& + & ducu - udcu).
\end{eqnarray}

In the present model it is useful 
to rewrite the above wave functions in terms of products of  symmetric 
$\phi_{[2]}(q_a q_b) = (q_a q_b + q_b q_a)/\sqrt{2} $ or  antisymmetric  
$\phi_{[11]}(q_a q_b) = (q_a q_b - q_b q_a)/\sqrt{2} $ 
quark pair states for the pairs 12 and 34.  This allows a forward 
calculation of the flavor integrated matrix elements (\ref{twobody}) and in addition one 
can easily read off the isospin of the corresponding wave function.

In particular the states (\ref{22s}) and (\ref{22a}) become

\begin{eqnarray}\label{22sprime}
|[22]_F 2211 \rangle &=& \sqrt{\frac{1}{6}} [ \sqrt{2}\phi_{[2]}(uu)~\phi_{[2]}(cd)\nonumber \\ 
& + &\sqrt{2}\phi_{[2]}(cd)~\phi_{[2]}(uu) \nonumber \\ 
& - & \phi_{[2]}(ud)~\phi_{[2]}(uc)  -  \phi_{[2]}(uc)~\phi_{[2]}(ud) ]
\end{eqnarray}
and 
\begin{eqnarray}\label{22aprime}
|[22]_F 2121 \rangle  &=& \sqrt{\frac{1}{2}}[ \phi_{[11]}(ud)~\phi_{[11]}(uc)\nonumber \\ 
& + & \phi_{[11]}(uc)~\phi_{[11]}(ud)],
\end{eqnarray}
where (\ref{22aprime}) obviously has isospin $I = 1/2$
which means that the pairs 12 and 34  in (\ref{22sprime})  have to couple to the same isospin value  as well.

For the irrep $[31]$ one has to use the Rutherford-Young-Yamanouchi basis, which symmetrizes or
antisymmetrizes the last two particles  \cite{Stancu:1991rc}. Accordingly
the  basis vectors  (\ref{31s}) and  (\ref{31m})  transform into
\begin{eqnarray}\label{31sprime} 
  | [31]_F {\overline {12}11} \rangle  &=& \sqrt{\frac{1}{6}}[\phi_{[2]}(uu)~\phi_{[2]}(cd) + \sqrt{2}\phi_{[2]}(ud)~\phi_{[2]}(uc)\nonumber \\
& - & \sqrt{2}\phi_{[2]}(uc)~\phi_{[2]}(ud) \nonumber \\
& - & \phi_{[2]}(cd)~ \phi_{[2]}(uu) ],
\end{eqnarray}
where the pair 34 is in a symmetric state, and
\begin{eqnarray}\label{31aprime}
| [31]_F {\tilde {12}11} \rangle  &=&\sqrt{\frac{1}{3}}[\phi_{[2]}(uu)~\phi_{[11]}(cd) \nonumber \\
& + & \sqrt{2}\phi_{[2]}(ud)~\phi_{[11]}(uc)],
\end{eqnarray}
where the pair 34 is in an antisymmetric state.

The state (\ref{31a}) can simply be rewritten as 
\begin{eqnarray}\label{31saprime}
| [31]_F 1121 \rangle  &=&\sqrt{\frac{1}{3}}[\phi_{[11]}(dc)~\phi_{[2]}(uu) \nonumber \\
& + & \sqrt{2}\phi_{[11]}(uc)~\phi_{[2]}(ud)],
\end{eqnarray}
where the pair 12 is in an antisymmetric state and 34 in a symmetric state. The states (\ref{31sprime})-(\ref{31saprime})
can have $I$ = 1/2 or 3/2.

For the irrep $[211]$ the vector (\ref{211s}) can be directly rewritten as
\begin{eqnarray}\label{211sameprime}  
| [211]_F 3211 \rangle  &=&  \frac{1}{2} (\sqrt{2} \phi_{[2]}(uu)~\phi_{[11]}(dc) - \phi_{[2]}(ud)~\phi_{[11]}(uc) \nonumber \\
& + &  \phi_{[2]}(uc)~\phi_{[11]}(ud)),
\end{eqnarray}
and the mixture of the other two in the Rutherford-Young-Yamanouchi basis reads
\begin{eqnarray}\label{211sprime} 
  | [211]_F {\overline {13}21} \rangle  &=& \frac{1}{2}[\phi_{[11]}(ud)~\phi_{[2]}(uc) + \sqrt{2}\phi_{[11]}(dc)~\phi_{[2]}(uu)\nonumber \\
& - & \phi_{[11]}(uc)~\phi_{[2]}(ud) ],
\end{eqnarray}
\begin{eqnarray}\label{211aprime}
| [211]_F {\tilde {13}21} \rangle  &=&\sqrt{\frac{1}{2}}[- \phi_{[11]}(ud)~\phi_{[11]}(uc) \nonumber \\
& + & \phi_{[11]}(uc)~\phi_{[11]}(ud)].
\end{eqnarray}
The states (\ref{211sameprime})-(\ref{211aprime}) have $I$ = 1/2.

\vskip 10pt
\begin{acknowledgement} 
I thank Ileana Guiasu for stimulating discussion.
This work has been supported by the Fonds de la Recherche Scientifique - FNRS, Belgium, 
under the grant number 4.4503.19.
\end{acknowledgement}



\begin{thebibliography}{9}
\bibitem{HO}
In a harmonic oscillator the kinetic energy is $KE = 
\frac{1}{2} \left[ N+3/2(A-1) \rangle \right]  \hbar \omega $,
where $N$ is the number of quanta and $A$ the number of particle.
For an excited pentaquark where $A$ = 5 and $N$ = 1 one has 
$KE = 7/2\hbar \omega$.  For a nucleon in the ground state 
one has  $KE = 3/2\hbar \omega$ and for a meson 
3/4 $\hbar \omega$, so that $\Delta KE =5/4 \hbar \omega$.
\end{thebibliography}
\end{document}